\documentclass[twocolumn,aps,pra,amssymb,showpacs,longbibliography,10pt]{revtex4-1}

\usepackage{amsfonts,amssymb,amsmath,graphicx,color,hyperref}

\newcommand{\br}{{\bf r}}
\newcommand{\vq}{{\bf r}}
\newcommand{\vj}{{\bf j}}

\newcommand{\imag}{\mathrm{Im}}
 
\definecolor{greenPR}{rgb}{0.00, 0.6, 0.00}
\newcommand{\commentout}[1]{}

\begin{document}

\title{Towards a quantum time mirror for nonrelativistic wave packets}

\author{Phillipp Reck$^1$, Cosimo Gorini$^1$, Arseni Goussev$^2$, Viktor Krueckl$^1$, Mathias Fink$^3$, Klaus Richter$^{1}$}

\email[]{klaus.richter@ur.de}

\affiliation{$^1$Institut f\"ur Theoretische Physik, Universit\"at Regensburg, 93040 Regensburg, Germany}

\affiliation{$^2$Department of Mathematics, Physics and Electrical Engineering, Northumbria University, Newcastle upon Tyne, NE1 8ST, United Kingdom}

\affiliation{$^3$Institut Langevin, ESPCI, CNRS, PSL Research University, 1 rue Jussieu, 75005, Paris, France}

\date{\today}

\begin{abstract}
	We propose a method -- a quantum time mirror (QTM) -- for simulating a partial time-reversal of the free-space motion of a nonrelativistic quantum wave packet. The method is based on a short-time spatially-homogeneous perturbation to the wave packet dynamics, achieved by adding a nonlinear time-dependent term to the underlying Schr\"odinger equation. Numerical calculations, supporting our analytical considerations, demonstrate the effectiveness of the proposed QTM for generating a time-reversed echo image of initially localized matter-wave packets in one and two spatial dimensions. We also discuss possible experimental realizations of the proposed QTM.
\end{abstract}


\maketitle


\section{Introduction}
\label{sec_intro}

The question of how to invert the time evolution of a wave, classical or quantum, in an efficient and controllable way has both fundamental and practical importance. The fundamental aspect of the question is evident from its connection with the problem of unidirectionality of the arrow of time, conceived in a seminal 19th century debate between Loschmidt and Boltzmann \cite{loschmidt1876, boltzmann1877}. 
The practical importance is apparent from numerous applications in medicine, telecommunication, material analysis, and, more generally, wave control \cite{fink1997, Fin99Time, LRTF07Focusing, mosk2012, bacot2015}.

One fruitful approach to the time inversion of classical wave motion is based on the concept of a time-reversal mirror: an array of receiver-emitter antennas is used to first record an incident wavefront, originating say from a localized source, and then to rebroadcast a time-inverted copy of the recording, thus generating a wave that effectively propagates backward in time and refocuses at the source point. To date, time reversal mirrors have been successfully implemented with acoustic \cite{fink1992,draeger1997}, elastic \cite{fink1997}, electromagnetic \cite{lerosey2004}, and water waves \cite{przadka2012,chabchoub2014}.

The classical procedure at the heart of such time-reversal mirrors, i.e. a continuous measurement and a subsequent reinjection of the signal, cannot be directly applied to quantum systems. The fundamental obstacle here is that any measurement performed on a quantum system 
is bound to perturb the quantum state and consequently affects its time evolution. (A theoretical 
scenario in which a time-dependent wave function is measured, recorded and then ``played back'' by a perfect non-invasive detector-emitter has been analyzed in Refs.~\cite{pastawski2007,calvo2010}) 
An alternative approach to manipulate the propagation of waves
relies on non-adiabatic perturbations to the system dynamics, such as an instantaneous change of its boundary conditions \cite{moshinsky1952,gerasimov1976,caslav1997,mendonca2002,delcampo2009,goussev2012a,haslinger2013,Gou13Diffraction}.
Protocols of this kind were considered for time- and space-modulated one-dimensional photonic \cite{sivan2011a,
sivan2011b, sivan2011c} and magnonic crystals \cite{chumak2010,karenowska2012}. 
More recently, Bacot {\it et al.}~ put forward and experimentally realized an instantaneous time mirror 
for gravity-capillary waves, requiring a sudden but homogenous modulation of water wave celerity \cite{bacot2015}.
Such approaches bypass the recording procedure, and are thus very appealing for quantum systems.  
A specific time-reversal protocol, albeit valid in a very narrow momentum range, 
was indeed devised for a one-dimensional periodically-kicked optical lattice \cite{martin2008} 
and realized in a $^{87}$Rb Bose-Einstein condensate (BEC) \cite{ullah2011}.
On the other hand, an instantaneous quantum time mirror (QTM) for Dirac-like systems (exploiting their spinor structure) was recently proposed in Ref.~\cite{reck2017}.

In this paper, we propose and investigate an experimentally realizable method for simulating the time-reversal of the free-space motion, thus mimicking a nonlinear QTM, for a spatially extended (orbital) quantum-mechanical wave function, such as a BEC cloud. The approach relies on generating a short-time spatially-uniform perturbation to the wave packet dynamics that corresponds to an additional nonlinear term added in the Schr\"odinger equation; in a BEC cloud system, such a perturbation can be realized using established experimental techniques allowing to tune the strength of the interaction among the cloud atoms \cite{inouye1998, roberts1998}.
More precisely, our QTM protocol comprises three stages: (i) a matter-wave packet propagates freely  in space for a time $0 < t < t_0$; (ii) at $t = t_0$, a strong nonlinear perturbation is switched on globally for a short period $\delta t$, leading to a near-instantaneous modification of the position-dependent phase of the wave function; (iii) the perturbation is switched off, the wave function evolves freely again, and at a time $t_{\rm echo} > t_0 + \delta t$ an ``echo'' signal of the original wave packet is observed.
From the conceptual viewpoint, the nonlinear QTM proposed in this paper can be viewed as a quantum-mechanical counterpart of the instantaneous time mirror for gravity-capillary waves by Bacot {\it et al.}~\cite{bacot2015}.

The paper is organized as follows. In Sec.~\ref{subsec_condensate}, we describe the physical principle underling the proposed QTM. Two concrete scenarios for generating a time-reversed motion of matter waves in one and two spatial dimensions are analyzed in Sec.~\ref{sec:examples}. A summary, concluding remarks, and a discussion of the feasibility of an experimental realization of the proposed QTM are presented in Sec.~\ref{sec_conclusion}. Technical calculations are deferred to an appendix.


\section{Physical principle of a nonlinear quantum time mirror}
\label{subsec_condensate}

We address the time evolution of a matter-wave packet $\Psi(\br,t)$, subject to the initial condition $\Psi(\br,0) = \Psi_0(\br)$, in accordance with the $(D+1)$-dimensional nonlinear Schr\"odinger equation 
\begin{equation}
i \hbar
\frac{\partial \Psi}{\partial t} = -\frac{\hbar^2}{2 m} \nabla^2 \Psi
+ \lambda f(t-t_0) |\Psi|^2 \Psi. 
\end{equation}
Here, $m$ is the atomic mass,
$\lambda$ quantifies the nonlinearity strength, and $t_0$ denotes the
time around which the nonlinear term representing interaction effects is switched on. The function $f(\zeta)$ is
sharply peaked around $\zeta\!=\!0$ and is chosen to satisfy the
normalization condition $\int_{-\infty}^{+\infty} \,{\rm d}\zeta f(\zeta) =
1$. We take $f$ to be a $\delta$-function in our analytical
calculations and a Gaussian peak $f(\zeta) = \frac{1}{\sqrt{2\pi}\Delta t} e^{-\frac{\zeta^2}{2\Delta t^2}}$ in all numerical simulations. The pulse length $\Delta t \ll t_0$ will be taken as $0.001 t_0$ and $0.0025t_0$ in one and two dimensions, respectively.

Wave packet dynamics in the presence of an infinitesimally short
nonlinear kick, $f(\zeta)\!=\!\delta(\zeta)$, can be described as follows. Rescaling $t \rightarrow t_0 t$, $\br \rightarrow
\sqrt{\frac{\hbar t_0}{m}} \br$, $\Psi \rightarrow \left(
\frac{m}{\hbar t_0} \right)^{D/4} \Psi$, and $\lambda \rightarrow
\hbar \left( \frac{\hbar t_0}{m} \right)^{D/2} \lambda$, we write the
nonlinear Schr\"odinger equation in a dimensionless form as
\begin{equation}
  i \frac{\partial \Psi}{\partial t} = -\frac{1}{2} \nabla^2 \Psi +
  \lambda \delta(t-1) |\Psi|^2 \Psi \,.
\label{NLSE}
\end{equation}
The evolution of the wave function from $\Psi_0(\br)$ at $t=0$ to its
value $\Psi_-(\br) = \Psi(\br,t=1^-)$ right before the kick is given by
\begin{equation}
  \Psi_-(\br) = \int \,{\rm d}^D\br' K(\br-\br',t)
  \Psi_0(\br') \,,
\label{free_prop}
\end{equation}
where the integration runs over the infinite $D$-dimensional space, and
 $  K({\bf q},t) \! = \! (2 \pi i t)^{-D/2} \exp{[i |{\bf q}|^2 / (2 t)]} $
is the free-particle propagator. The nonlinear kick results in an
instantaneous change of the wave function from $\Psi_-(\br)$ at $t=1^-$
to
\begin{equation}
  \Psi_+(\br) = \Psi_-(\br) e^{-i \lambda |\Psi_-(\br)|^2}
\label{Psi_jump}
\end{equation}
at $t=1^+$. Indeed, during the time interval $1^- < t < 1^+$, the wave
function transformation is dominated by the second term in the
right-hand side of Eq.~(\ref{NLSE}), and effectively governed by the
differential equation $\frac{\partial \ln \Psi(\br,t)}{\partial t} =
-i \lambda |\Psi_-(\br)|^2 \delta(t-1)$, the solution of which is
given by Eq.~(\ref{Psi_jump}).  After the kick, the wave function
evolves freely, so that $\Psi(\br,t) = \int \,{\rm d}^D \br' K(\br-\br',t)
\Psi_+(\br')$ for all times $t > 1$.

As evident from Eq.~(\ref{Psi_jump}), the instantaneous nonlinear kick
alters the phase of the wave function without producing any
probability density redistribution, so that $\rho(\br) \equiv
|\Psi_+(\br)|^2 = |\Psi_-(\br)|^2$. The phase change however affects
the probability current, whose dimensionless expression reads
$\vj(\br,t) = \imag \big[ \Psi^*(\br,t) \nabla \Psi(\br,t) \big]$. A straightforward
evaluation of the current right after the kick, $\vj_+ = \imag \big[
  \Psi_+^* \nabla \Psi_+ \big]$, yields
\begin{equation}
  \vj_+ = \vj_- + \Delta \vj \qquad \mathrm{with} \qquad \Delta \vj =
  -\lambda \rho \nabla \rho \,,
\label{current}
\end{equation}
where $\vj_- = \imag \big[ \Psi_-^* \nabla \Psi_- \big]$ is the
probability current immediately preceding the kick.  This in turn means
that, by properly tuning the kicking strength $\lambda$, the wave
propagation direction can be reversed for those parts of the matter
wave for which the vector $\nabla \rho$ is aligned (or anti-aligned)
with $\vj_-$.  Below we show that in geometries
accessible in atom-optics experiments this reversal effect is robust
and well-pronounced.

\section{The quantum time mirror at work}
\label{sec:examples}

\begin{figure*}[h!]
	\includegraphics[width=0.9\textwidth]{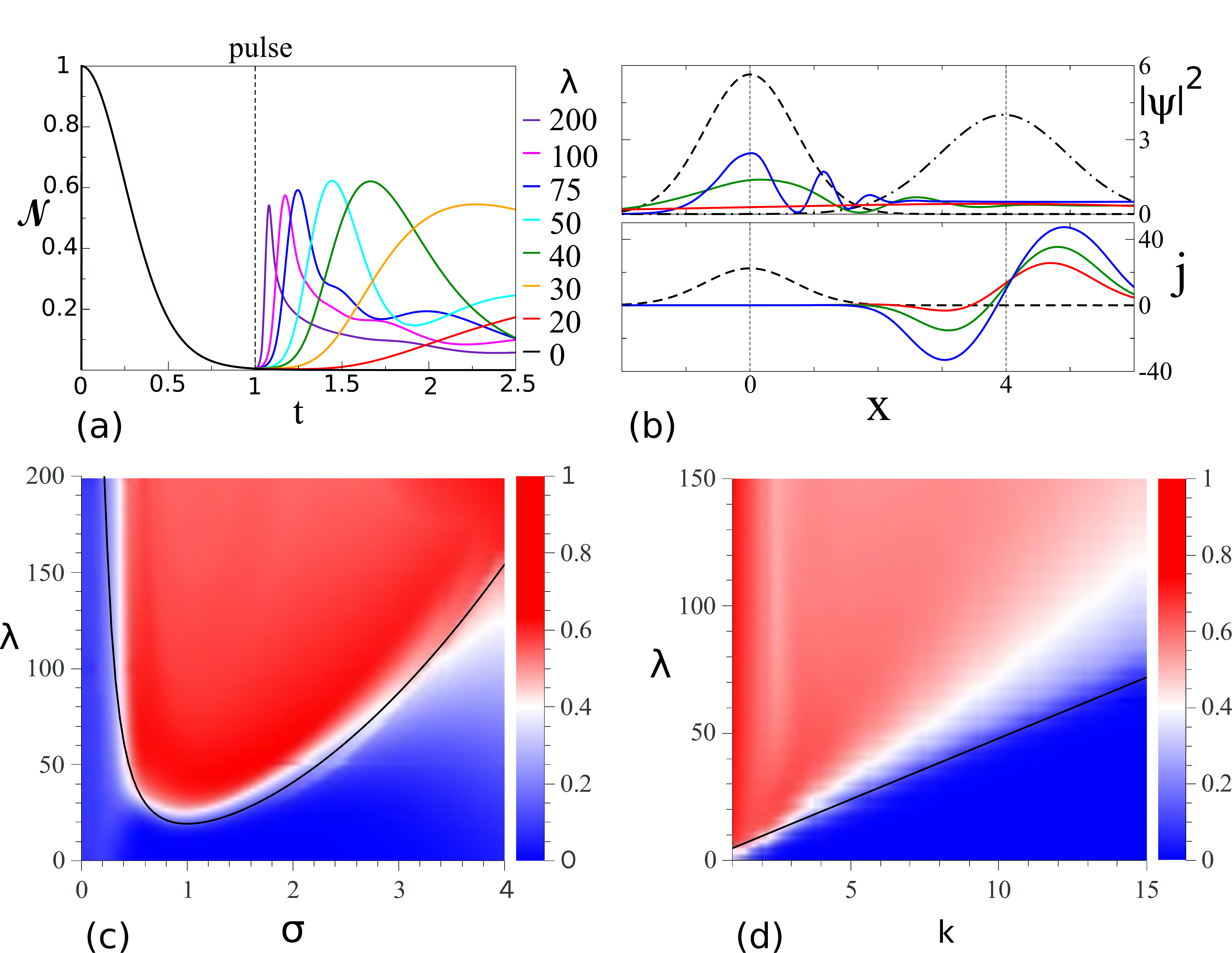}
	\caption{Echo of a 1D Gaussian wave packet subjected to a short, nonlinear pulse. All quantities are dimensionless according to the rescaling in the text above Eq.~\eqref{NLSE}.  (a) Norm correlation, Eq.~\eqref{eq:normcorr_ref}, as a function of time for various pulse strengths $\lambda$ of a kick at $t=1$ at fixed wave packet width $\sigma = 1$ and momentum $k = 4$, yielding echoes up to $60\%$. (b) Upper panel: real space probability density $\rho = |\psi|^2$ at $t=0$ (black dashed curve), $t=0.99$ (just before pulse, black dashed-dotted curve), and at the peak echo times (colored curves, color code as in (a)).  Notice that the latter depend on $\lambda$, as evident from (a). Lower panel: current density $j$ at $t=0$ (black dashed curve) and $t=1.01$ (right after the pulse, color code as in (a)). A negative current density indicates the part of the wave packet reversing its propagation direction thereby causing the echo. The minimal kicking strength for the used parameters, as predicted by Eq.~\eqref{eq:l_min1D}, is $\lambda_{\min} \simeq 20$ (red curve); the associated negative current density is not sufficient for echo generation. Panels (c) and (d) show the echo strength, i.e.\ the maximal achieved value of the norm correlation after the pulse for a given parameter set. The parameter, which is kept constant is (c) $k=4$, (d) $\sigma = 1$. The black curves are the analytical approximations (Eq.~\eqref{eq:l_min1D} for the minimal pulse strength \(\lambda_{\min}\) required to generate an echo and matches well the numerical results.}
\label{fig:1dNLSE}
\end{figure*}

As our first example, we consider the case of the initial state given
by a 1D Gaussian wave packet,
\begin{equation}
  \Psi_0(x) = (\pi \sigma^2)^{-1/4} \exp \left( -\frac{x^2}{2
    \sigma^2} + i k x \right),
\label{eq:wp_1d}
\end{equation}
characterized by the dimensionless real spatial dispersion $\sigma$
and average momentum $k$. The corresponding wave function at time
$t=1^-$ is obtained from Eq.~(\ref{free_prop}) and reads (up to a
position-independent phase factor) $\Psi_-(x) = [ \pi (\sigma^2 +
  \sigma^{-2}) ]^{-1/4} \exp [ -\xi^2 / 2 (\sigma^2 + i) + i k \xi ]$
with $\xi = x - k$ denoting the distance from the wave packet
center. Thus, the probability density at $t=1$ is $\rho = \exp(-\xi^2
/ \sigma_1^2) / \sqrt{\pi} \sigma_1$, where $\sigma_1 = \sqrt{\sigma^2
  + \sigma^{-2}}$ is the dispersion of the wave packet at the time of
the kick. The probability currents before and after the kick are,
respectively, $j_- = \left( k + \frac{\xi}{\sigma^2 \sigma_1^2}
\right) \rho$ and $j_+ = j_- + \Delta j$, with $\Delta j = \frac{2
  \lambda \xi}{\sigma_1^2} \rho^2$. The minimal kick strength
$\lambda_{\min}$ necessary for reversing the direction of
motion of (and effectively reflecting) a part of the wave packet can
be estimated by requiring $j_+ = 0$ at $\xi = -\sigma_1$. In the case
of a fast moving wave packet, such that $k \gg 1 / \sigma^2 \sigma_1$,
this estimation yields
\begin{equation}
  \lambda_{\min} \simeq C k \left( \sigma^2 + \frac{1}{\sigma^2}
  \right) \label{eq:l_min1D}
\end{equation}
with $C = e \sqrt{\pi} / 2 \simeq 2.4$. Then, given a kicking strength
$\lambda > \lambda_{\min}$, the time $t_{\mathrm{echo}}$ at which the
reflected part of the wave packet reaches its initial position,
leading to a partial echo of the original wave packet, can be
evaluated as follows. The velocity of the reflected wave is
$k_{\mathrm{rev}} \simeq j_+ \big|_{\xi = -\sigma_1} \simeq k -
\frac{\lambda}{C \sigma_1^2} = \left( 1 -
\frac{\lambda}{\lambda_{\min}} \right) k$, and the revival occurs when
$|k_{\mathrm{rev}}| (t_{\text{echo}} - 1) = k$ or, correspondingly, at
\begin{equation}
  t_{\text{echo}} = \frac{\lambda}{\lambda - \lambda_{\min}} \,.
\label{t_echo_BEC}
\end{equation}

The numerical simulations are based on the wave packet propagation algorithm 
Time-dependent Quantum Transport (TQT) \cite{krueckl2013}. The state is discretized on a square grid and the time evolution is calculated for sufficiently small time steps such that the Hamilton operator can be assumed time independent for each step. We calculate the action of $H$ on $\psi$ in a mixed position and momentum-space representation by the application of Fourier Transforms. With this a Krylov Space is spanned, which can be used to calculate the time evolution using a Lanczos method \cite{lanczos1950}.

The echo strength is quantified by the norm correlation between the initial and the time propagated 
wave packet defined as \cite{eckhardt2003}
\begin{equation}
\label{eq:normcorr_ref}
 \mathcal{N}(t) = \frac{\int \,{\rm d}^D \br \left|\Psi_0(\br)\right|^2\left|\Psi(\br,t)\right|^2}{\sqrt{\int  \,{\rm d}^D \br \left|\Psi_0(\br)\right|^4\int  \,{\rm d}^D \br\left|\Psi(\br,t)\right|^4}}.
\end{equation}
Figure \ref{fig:1dNLSE}a) presents $\mathcal{N}(t)$ for various pulse strengths $\lambda$ at constant $\sigma$ and $k$, demonstrating echo strengths up to $60\%$. 
The occurring lower peaks at higher $\lambda$ for larger times are due secondary peaks of the distorted wave packet, 
which can be seen in Fig.~\ref{fig:1dNLSE}b): This panel shows the spatial probability density $\rho= |\psi|^2$
at times $t=0$ (black dashed curve) and $t=0.99$ (immediately before pulse, black dashed-dotted curve), 
as well as the reflected wave packets for different $\lambda$'s (color code as in panel (a)),
each shown at its peak echo time.
In the lower plot, the current density $j$ is shown directly after the pulse $t=1.01$.  Parts of the wave packet with negative current density move backwards leading to the echo. 
For the parameters used, the estimated value for the minimal pulse strength in (\ref{eq:l_min1D}) is $\lambda_{\min} \simeq 20$ corresponding to the red curve, whose current density exhibits only a vanishing negative part that is insufficient for echo generation, thus verifying the prediction \eqref{eq:l_min1D}.


To explore the parameter space for the possibility of achieving echoes, the peak of the norm correlation (in time) is plotted
as a function of $\lambda$ and $\sigma$ in Fig.~\ref{fig:1dNLSE}c) and as a function of $\lambda$ and $k$ 
 in Fig.~\ref{fig:1dNLSE}d).
The black curve shows the analytic approximation \eqref{eq:l_min1D} of the minimal pulse strength $\lambda_{\min}$. Although it does not fit perfectly, the analytic approximation is in good agreement and 
still well-suited to approximate the minimal pulse strength required for a time-reversal.

\begin{figure*}[h!]
	\includegraphics[width=0.5\textwidth]{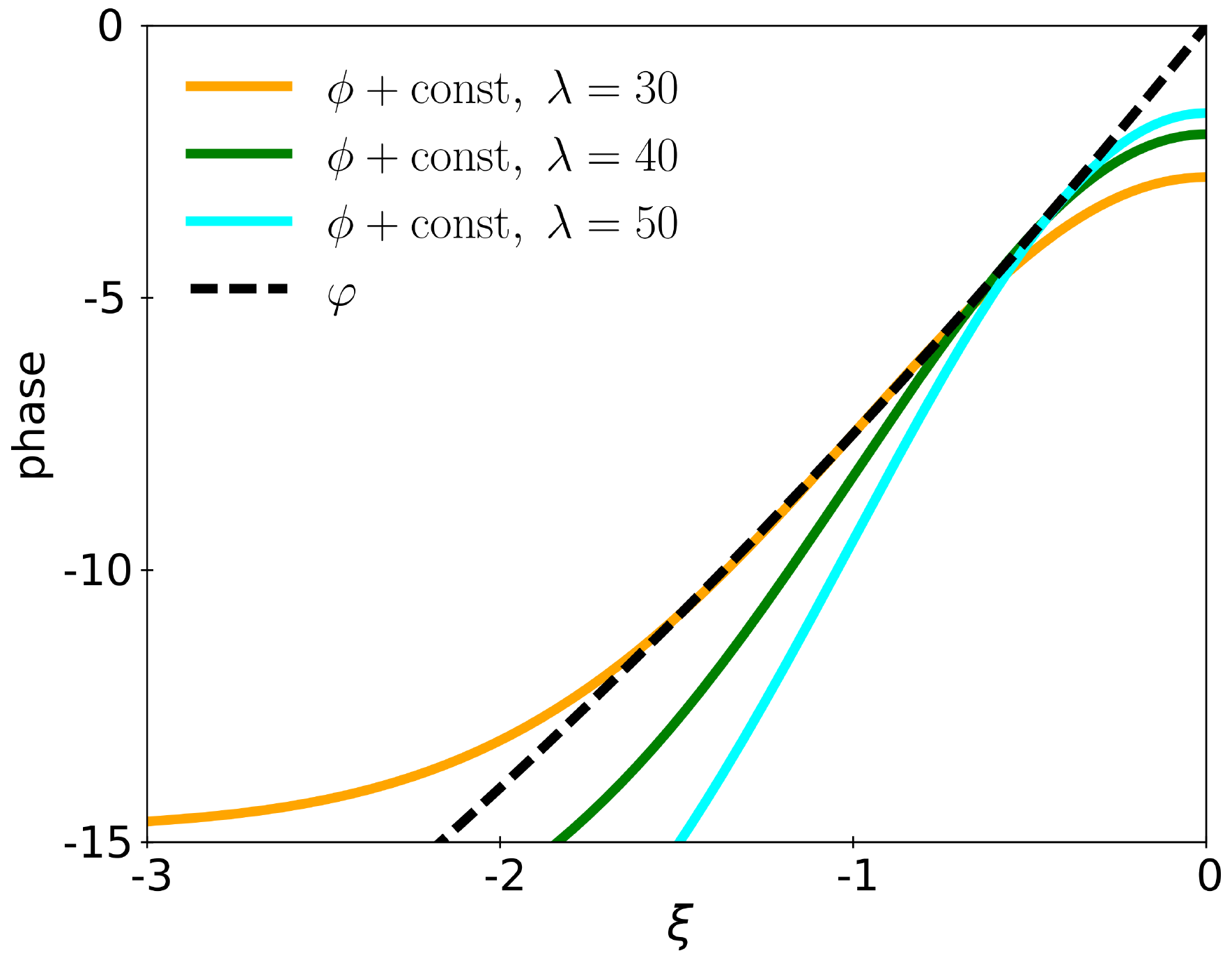}
	\caption{Comparison between the QTM-imprinted phase $\phi(\xi)$ (shifted by a physically irrelevant constant) and ideal-time-reversal phase $\varphi(\xi)$, as given by Eqs.~(\ref{phase:QTM}) and (\ref{phase:ideal}), respectively. The system parameters are the same as in Fig.~\ref{fig:1dNLSE}a): $\sigma = 1$ and $k = 4$. Three different values of the kicking strength are considered: $\lambda = 30$ (orange curve), $\lambda = 40$ (green curve) and $\lambda = 50$ (cyan curve). (The color code coincides with the one adopted in Fig.~\ref{fig:1dNLSE}a).)}
	\label{fig:1d_phases}
\end{figure*}

In order to better understand the quality, underlying principles and limitations of the proposed QTM, it is instructive to compare the state $\Psi_+$, rendered by the nonlinear kick (see Eq.~(\ref{Psi_jump})), against the desired (perfectly time-reversed) state $\mathcal{C} \Psi_-$, obtained by applying the (anti-unitary) complex conjugation operator $\mathcal{C}$ to the pre-kick state $\Psi_-$. In the case of the initial state given by Eq.~(\ref{eq:wp_1d}), we have
\begin{equation}
	\Psi_+ = \Psi_- e^{-i \phi} \,, \qquad \phi = \frac{\lambda}{\sqrt{\pi} \sigma_1} e^{-( \xi / \sigma_1 )^2}
\label{phase:QTM}
\end{equation}
and
\begin{equation}
	\mathcal{C} \Psi_- = \Psi_- e^{-i (\varphi + \varphi_0)} \,, \qquad \varphi = \left( \frac{\xi}{\sigma_1 \sigma} \right)^2 + 2 k \xi \,,
\label{phase:ideal}
\end{equation}
where $\sigma_1 = \sqrt{\sigma^2 + \sigma^{-2}}$ is the width of wave packet at time $t = 1$ when the kick occurs, $\xi = x - k$ is the distance measured from the center of the wave packet, and $\varphi_0$ is a constant (position-independent) phase related to the global phase of $\Psi_-$. While, in general, the two phases, $\phi$ and $\varphi$, have different functional forms, $\phi(\xi)$ may serve as a reasonable approximation to $\varphi(\xi)$, modulo a physically irrelevant constant shift, over a finite position interval. It is the probability density supported by this position interval that makes the main contribution to the time-reversed wave generated by the nonlinear QTM. Figure~\ref{fig:1d_phases} presents a comparison between the ideal (target) phase $\varphi(\xi)$ and the phase $\phi(\xi)$ imprinted by the proposed QTM. The system parameters are taken to be the same as in Fig.~\ref{fig:1dNLSE}a), i.e. $\sigma = 1$ and $k = 4$, and the three values of the kicking strength considered are $\lambda = 30$, $40$, and $50$.

\begin{figure*}[h!]
	\includegraphics[width=0.8\textwidth]{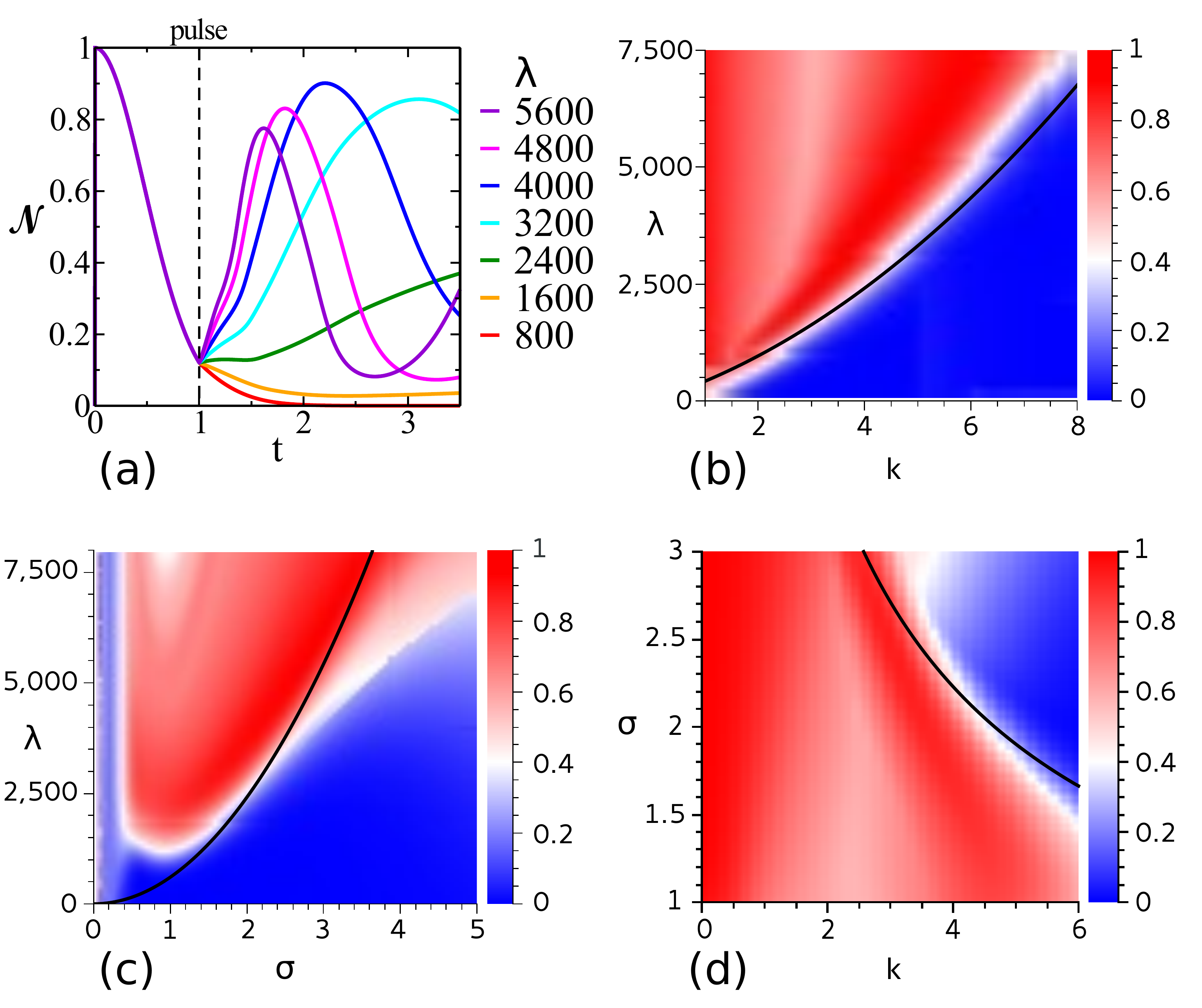}
	\caption{Echo of a 2D Gaussian wave packet subjected to a short, nonlinear pulse. (a) The norm correlation \eqref{eq:normcorr_ref} is shown as function of time for varying pulse strength $\lambda$ and fixed momentum $k = 4$ and width $\sigma = 2$. A norm correlation up to 90\% is achievable. Note that the large $\lambda$ values are due to the variable rescaling as described in the text. (b)-(d) The echo peak of the norm correlation is plotted as a function of $\lambda$, $\sigma$ and $k$. The constant parameters are (b) $\sigma =2$, (c) $k=4$ and (d) $\lambda = 3000$. The black curves are the analytical approximations by Eq. \eqref{eq:2dNLSE_lmin} for the minimal pulse strength \(\lambda_{\min}\) required to generate an echo. The radius of the Gaussian ring, is $R=6$.}
\label{fig:2D-NLSE}
\end{figure*}

We further investigate the dynamics of a 2D wave packet, initially given by
\begin{equation}
  \Psi_0(\br) = \sqrt{\frac{1}{2 \pi^{3/2} R \sigma}} \exp \left[
    -\frac{(r-R)^2}{2 \sigma^2} + i k r \right]
\label{eq:wp_2d}
\end{equation}
with $r = |\br|$ and $k > 0$. For $R \gg \sigma$, the wave function is normalized to one and describes a Gaussian ring of radius $R$ and width $\sigma$ that spreads radially with the average velocity $k$. A straightforward, although tedious, calculation shows that, in the parametric regime defined by $1 \ll \sigma \ll R$ and $kR \gg 1$, the wave packet at $t = 1^-$ has the form (up to a spatially uniform phase) (see Appendix~\ref{app:ring}):
\begin{equation}
	\Psi_-(\br) = \sqrt{\frac{1}{2 \pi^{3/2} R_1 \sigma}} \exp \left[ -\frac{(r-R_1)^2}{2 \sigma^2} + i k r \right] \,,
\label{eq:ring_wp}
\end{equation}
where $R_1 = R + k$ is the radius of the Gaussian ring at time $1^-$. Thus, the corresponding
probability density is given by $\rho = (2 \pi^{3/2} R_1 \sigma)^{-1}
\exp [ -(r-R_1)^2 / \sigma^2 ]$, and the probability current at
$t=1^+$, reads $\vj_+ = \left( k \rho + \frac{2 \lambda
  (r-R_1)}{\sigma^2} \rho^2 \right) \frac{\br}{r}$. Then, the
evaluation of the minimal kick strength required to trigger a
probability density echo proceeds in close analogy with the
corresponding 1D calculation, resulting in
\begin{equation}
  \lambda_{\min} \simeq 2 \pi C (R + k) k \sigma^2 \,.  \label{eq:2dNLSE_lmin}
\end{equation}
Finally, just as in the 1D case, the echo time is determined by Eq.~(\ref{t_echo_BEC}).

The numerical calculations (Fig.~\ref{fig:2D-NLSE}) attest the possibility of pronounced echoes up to 90\% also in the 2D setup. Although the parameter range is not in the regime of the analytical approximation, the value of $\lambda_{\min}$ in Eq.~(\ref{eq:2dNLSE_lmin}) is still well-suited to estimate the minimal pulse strength $\lambda$ required (see black curves). 

The color plots in Figs.~\ref{fig:1dNLSE} and \ref{fig:2D-NLSE} seem to imply $\lambda>\lambda_{\min}$ (marked as black lines) to be the only echo requirement for echo generation. 
However for large $\lambda$ the wave packet splits into many peaks, as shown in Fig. \ref{fig:1dNLSE}b), blue curve for $\lambda = 200$. 
In such a scenario the norm correlation is still fairly high, but the wave packet might not longer have the desired shape. 
The effect of many peaks, {\it i.e.} very large $\lambda$, on the norm correlation can be seen in Fig.~\ref{fig:2D-NLSE}c), 
where the echo strength moderately declines for $\sigma\approx 1$ and $\lambda > 3000$.



\section{Summary and conclusions}
\label{sec_conclusion}

In summary, we have proposed a protocol for simulating the time-reversed motion of a localized matter-wave packet evolving in free space.
Our method is based on making a near-instantaneous spatially-homogeneous perturbation to the wave packet dynamics by externally switching on a nonlinear perturbation for a short time interval. The analytical and numerical considerations presented in our paper demonstrate the efficiency of the proposed quantum time mirror in one and two spatial dimensions.

We note that the time reversal protocol proposed in this paper could in principle be employed in different physical systems, such as ultracold atomic clouds, optical pulses, or shallow water waves, as long as the system's time evolution is governed by the nonlinear Schr{\"o}dinger equation. Here, we further explore the possible connection between the numerical simulations reported in this paper and relevant atom-optics experiments. To this end, we provide an estimate for values of the dimensionless parameters $\sigma$ and $k$, defined in Eqs.~(\ref{eq:wp_1d}) and (\ref{eq:wp_2d}), for the case of ultracold lithium atoms. The mass of a $^7$Li atom is $m = 7.016 \, \mathrm{u} = 1.165 \times 10^{-26} \, \mathrm{kg}$. Taking the wave packet propagation time until the nonlinear kick to be $t_0 = 10 \, \mathrm{ms}$, we see that the wave packet width range of $10-50 \, \mu\mathrm{m}$ corresponds to $1.05 < \sigma < 5.26$, and the mean velocity range of $2-10 \, \mathrm{mm\,s^{-1}}$ corresponds to $2.1 < k < 10.5$. These parameter ranges coincide with the ones considered in this paper, which strongly suggests that the matter wave reversal effects predicted here can be realized in experiments with lithium BECs.

In order to further facilitate experimental realization of the proposed QTM, we make a rough estimate of the scattering length of condensed lithium atoms required to generate a reflected wave. In the one-dimensional case, the (dimensional) kicking strength $\lambda \hbar \sqrt{\hbar t_0 / m} / \Delta t$ (see the discussion preceding Eq.~(\ref{NLSE})) is approximately equal to $2 N \hbar^2 a_s / (m a_{\perp}^2)$, where $N$ is the number of condensed atoms, $a_s$ is the scattering length, $\Delta t$ is the kick duration (taken to be $\Delta t = 0.001 t_0 = 10 \, \mu\mathrm{s}$ in our numerical simulations), and $a_{\perp}$ is the linear length scale of the potential confining the atomic motion in the transverse direction \cite{SPR02Effective}. This yields the estimate $a_s = \lambda a_{\perp}^2 \sqrt{m t_0 / \hbar} / (2 N \Delta t)$. Taking $N = 10^7$, $a_{\perp} = 10 \, \mu\mathrm{m}$, and $\lambda$ in the range $10-200$ (see, e.g., Fig.~\ref{fig:1dNLSE}), we find the required scattering length to lie in the range $5 \, n\mathrm{m} \lesssim a_s \lesssim 105 \, n\mathrm{m}$. While challenging, the suggested parameter values are not impossible to achieve in modern atom-optics experiments, using, for instance, such novel techniques as optical control of Feshbach resonances \cite{CHXC15Quantum}.


\begin{acknowledgments}
	The authors thank Ilya Arakelyan for useful discussions. A.G. acknowledges the support of EPSRC Grant No.~EP/K024116/1. C.G., V.K., K.R. and P.R. acknowledge support from Deutsche Forschungsgemeinschaft within SFB~689 and GRK~1570.
\end{acknowledgments}

\appendix

\onecolumngrid

\section{Free spreading of a Gaussian ring wave packet in 2D: Derivation of Eq.~(\ref{eq:ring_wp})}
\label{app:ring}

Let us consider a two-dimensional wave packet, initially (at $t=t_0$)
given by
\begin{equation*}
	\Psi_0(\vq) = C \exp \left( -\frac{(r-R)^2}{2 \sigma^2} + i k_0
	(r-R) \right)
\end{equation*}
with $r = |\vq| = \sqrt{x^2 + y^2}$, $\sigma \ll R$, $k_0 >
0$ and $C \simeq 1 / \sqrt{2 \pi^{3/2} R \sigma}$, so that the probability density is normalized to unity. Let $\Psi(\vq,t)$ be the wave packet evolved from $\Psi_0(\vq)$ in the course of a free-particle evolution through time $t$. Here, we would like to show that in the parametric regime given by $\sqrt{\hbar t / m} \ll \sigma \ll R$ and $k_0 R \gg 1$ the wave packet $\Psi$ has the same functional dependence on $\vq$ as $\Psi_0$.

The free-particle propagator in 2D reads
\begin{equation*}
	K_0(\vq, \vq', t) = \frac{m}{2 \pi i \hbar t} \exp \left( i \frac{m
		(\vq-\vq')^2}{2 \hbar t} \right) \,.
\end{equation*}
Hence,
\begin{align*}
	\Psi(\vq,t) &= \frac{m C}{2 \pi i \hbar t} \int\,{\rm d}^2 \vq' \exp \left(
	-\frac{(r'-R)^2}{2 \sigma^2} + i k_0 (r'-R) + i \frac{m
		(\vq-\vq')^2}{2 \hbar t} \right) \nonumber\\ &= \frac{m C}{2 \pi i
		\hbar t} \int_0^{\infty} \,{\rm d} r' r' \int_0^{2 \pi} \,{\rm d} \theta \, \exp
	\left( -\frac{(r'-R)^2}{2 \sigma^2} + i k_0 (r'-R) + i \frac{m (r^2
		+ r'^2 - 2 r r' \cos\theta)}{2 \hbar t} \right) \nonumber\\ &=
	\frac{m C}{i \hbar t} \exp \left( -\frac{R^2}{2 \sigma^2} - i k_0 R
	+ i \frac{m r^2}{2 \hbar t} \right) G(r, t) \,,
\end{align*}
where
\begin{equation*}
	G(r, t) = \int_0^{\infty} \,{\rm d} r' r' J_0\left( \frac{m r r'}{\hbar t}
	\right) \exp \left[ -\frac{1}{2} \left( \frac{1}{\sigma^2} - i
	\frac{m}{\hbar t} \right) r'^2 + \left( \frac{R}{\sigma^2} + i k_0
	\right) r' \right] \,.
\end{equation*}

Let us investigate the behavior of $G(r,t)$ around the spatial point
$R + \frac{\hbar k_0}{m} t$. Taking into account the fact that the
main contribution to the integral comes from the region $|r' - R|
\lesssim \sigma$, we have
\begin{equation*}
	\frac{m r r'}{\hbar t} \sim \frac{m R}{\hbar t} \left( R +
	\frac{\hbar k_0}{m} t \right) = \frac{m R^2}{\hbar t} + k_0 R > k_0
	R \,.
\end{equation*}
Assuming $k_0 R \gg 1$, we see that the argument of the
Bessel function is always large compared to one, {\it i.e.}\, $\frac{m r
	r'}{\hbar t} \gg 1$. This allows us to use the large argument
asymptotics
\begin{equation*}
	J_0 \left( \frac{m r r'}{\hbar t} \right) \simeq \sqrt{\frac{2}{\pi
			\frac{m r r'}{\hbar t}}} \cos \left( \frac{m r r'}{\hbar t} -
	\frac{\pi}{4} \right) = \sqrt{\frac{\hbar t}{2 \pi m r r'}} \left[
	e^{i \left( \frac{m r r'}{\hbar t} - \frac{\pi}{4} \right)} +
	e^{-i \left( \frac{m r r'}{\hbar t} - \frac{\pi}{4} \right)}
	\right]
\end{equation*}
to write
\begin{equation*}
	\Psi(\vq, t) \simeq \frac{C}{i} \sqrt{\frac{m R}{2 \pi \hbar t r}}
	\exp \left( -\frac{R^2}{2 \sigma^2} - i k_0 R + i \frac{m r^2}{2
		\hbar t} \right) \left[ \Phi_+(r, t) + \Phi_-(r, t) \right] \,.
\end{equation*}
Here,
\begin{align*}
	\Phi_{\pm} (r, t) &= e^{\mp i \frac{\pi}{4}}
	\int_{-\infty}^{+\infty} \,{\rm d} r' \sqrt{\frac{r'}{R}} \exp \left\{
	-\frac{1}{2} \left( \frac{1}{\sigma^2} - i \frac{m}{\hbar t} \right)
	r'^2 + \left[ \frac{R}{\sigma^2} + i \left( k_0 \pm \frac{m r}{\hbar
		t} \right) \right] r' \right\} \nonumber\\ &\simeq e^{\mp i
		\frac{\pi}{4}} \int_{-\infty}^{+\infty} \,{\rm d} r' \exp \left\{
	-\frac{1}{2} \left( \frac{1}{\sigma^2} - i \frac{m}{\hbar t} \right)
	r'^2 + \left[ \frac{R}{\sigma^2} + i \left( k_0 \pm \frac{m r}{\hbar
		t} \right) \right] r' \right\} \nonumber\\ &= e^{\mp i
		\frac{\pi}{4}} \sqrt{\frac{2 \pi}{\frac{1}{\sigma^2} - i
			\frac{m}{\hbar t}}} \exp \frac{\left[ \frac{R}{\sigma^2} + i
		\left( k_0 \pm \frac{m r}{\hbar t} \right) \right]^2}{2 \left(
		\frac{1}{\sigma^2} - i \frac{m}{\hbar t} \right)} \,.
\end{align*}
Introducing
\begin{equation*}
	\epsilon = \frac{\hbar t}{m \sigma^2} \quad \mathrm{and} \quad v_0 =
	\frac{\hbar k_0}{m} \,,
\end{equation*}
we rewrite the previous expression as
\begin{equation*}
	\Phi_{\pm} (r, t) = e^{\mp i \frac{\pi}{4}} \sqrt{\frac{2 \pi i
			\hbar t}{m (1 + i \epsilon)}} \exp \left( i \frac{m}{2 \hbar t}
	\frac{[\epsilon R + i (v_0 t \pm r)]^2}{1 + i \epsilon} \right) \,.
\end{equation*}
This leads to
\begin{align*}
	\Psi(\vq, t) &= \frac{C}{i} \sqrt{\frac{m R}{2 \pi \hbar t r}} \exp
	\left( \frac{m}{2 \hbar t} [-\epsilon R^2 - 2 i R v_0 t + i r^2]
	\right) \left[ \Phi_+(r, t) + \Phi_-(r,t) \right] \nonumber\\ &= C
	\sqrt{\frac{R}{i (1 + i \epsilon) r}} \sum_{\gamma = \pm 1} e^{-i
		\gamma \frac{\pi}{4}} \exp \left\{ \frac{m}{2 \hbar t} \left(
	-\epsilon R^2 - 2 i R v_0 t + i r^2 + i \frac{[\epsilon R + i (v_0 t
		+ \gamma r)]^2}{1 + i \epsilon} \right) \right\} \nonumber\\ &=
	C \sqrt{\frac{R}{i (1 + i \epsilon) r}} \sum_{\gamma = \pm 1} e^{-i
		\gamma \frac{\pi}{4}} \exp \left( -\frac{m}{2 \hbar t} \,
	\frac{\epsilon (R + \gamma r)^2 + i \big[ 2 (R + \gamma r) + v_0 t
		\big] v_0 t}{1 + i \epsilon} \, \frac{1 - i \epsilon}{1 - i
		\epsilon}\right) \nonumber\\ &= C \sqrt{\frac{R}{i (1 + i
			\epsilon) r}} \sum_{\gamma = \pm 1} e^{-i \gamma \frac{\pi}{4}}
	\exp \left( -\frac{m}{2 \hbar t} \, \frac{\epsilon (R + v_0 t +
		\gamma r)^2 + i 2 (R + v_0 t + \gamma r) v_0 t - i \big[ (v_0 t)^2
		+ \epsilon^2 (R + \gamma r)^2\big]}{1 + \epsilon^2} \right)\, .
\end{align*}
Assuming further that $\epsilon \ll 1$, we have
\begin{align*}
	\Psi(\vq, t) &\simeq C \sqrt{\frac{R}{i r}} \sum_{\gamma = \pm 1}
	e^{-i \gamma \frac{\pi}{4}} \exp \left\{ -\frac{m}{2 \hbar t} \left[
	\epsilon (R + v_0 t + \gamma r)^2 + i 2 (R + v_0 t + \gamma r) v_0
	t - i (v_0 t)^2 \right] \right\} \nonumber\\ &= C \sqrt{\frac{R}{i
			r}} \sum_{\gamma = \pm 1} e^{-i \gamma \frac{\pi}{4}} \exp
	\left( -\frac{(R + v_0 t + \gamma r)^2}{2 \sigma^2} - i k_0 (R + v_0
	t + \gamma r) + i \frac{\hbar k_0^2 t}{2 m} \right) \,.
\end{align*}
Taking into account that last expression for $\Psi$ is only valid for
$r$ close to $r_t$, where
\begin{equation*}
	r_t = R + v_0 t = R + \frac{\hbar k_0 t}{m} \,,
\end{equation*}
we see that a contribution of the term with $\gamma = +1$ is
negligibly small. Thus we arrive at the following expression for the freely propagated wave function:
\begin{equation*}
	\Psi(\vq, t) \simeq \sqrt{\frac{1}{2 \pi^{3/2} \sigma r_t}}
	\exp \left( -\frac{(r - r_t)^2}{2 \sigma^2} + i k_0 (r - r_t) + i
	\frac{\hbar k_0^2 t}{2 m} \right) \,.
\end{equation*}
This expression is only valid in the parametric regime defined by
\begin{equation*}
	\sqrt{\frac{\hbar t}{m}} \ll \sigma \ll R \quad \mathrm{and} \quad k_0 R \gg 1 \,.
\end{equation*}

\twocolumngrid

\bibliography{QTMs_biblio}


\end{document}